\documentclass[runningheads]{llncs}
\usepackage[T1]{fontenc}
\usepackage{graphicx}
\usepackage{tikz}
\usepackage{amsmath}
\usepackage{booktabs}
\usepackage{multirow}
\usepackage{multicol}
\usepackage{algpseudocode}
\usepackage{algorithm}

\algrenewcommand\algorithmicrequire{\textbf{Input:}}
\algrenewcommand\algorithmicensure{\textbf{Output:}}
\begin{document}
\title{S3PHER: Secure and Searchable System for Patient-driven HEalth data shaRing}
\titlerunning{S3PHER}
%
\author{Ivan Costa\inst{1,3}\orcidID{0009-0009-8480-9352} \and
Ivone Amorim\inst{2}\orcidID{0000-0001-6102-6165} \and
Eva Maia\inst{1}\orcidID{0000-0002-8075-531X} \and Pedro Barbosa\inst{1}\orcidID{0000-0002-7381-3462} \and Isabel Praça\inst{1}\orcidID{0000-0002-2519-9859}}
\authorrunning{I. Costa et al.}
%
\institute{GECAD - Research Group on Intelligent Engineering and Computing for Advanced Innovation and Development, Porto, Portugal \and
IPP - Instituto Politécnico do Porto, Porto, Portugal \and \email{ivcsi@isep.ipp.pt}
}
\maketitle              
\begin{abstract}
Healthcare data contains some of the most sensitive information about an individual, yet sharing this data with healthcare practitioners can significantly enhance patient care and support research efforts. However, current systems for sharing health data between patients and caregivers do not fully address the critical security requirements of privacy, confidentiality, and consent management. Furthermore, compliance with regulatory laws such as GDPR and HIPAA is often deficient, largely because patients typically are asked to provide general consent for healthcare entities to access their data.
Recognizing the limitations of existing systems, we present S3PHER, a novel approach to sharing health data that provides patients with control over who accesses their data, what data is accessed, and when. Our system ensures end-to-end privacy by integrating a Proxy Re-Encryption Scheme with a Searchable Encryption Scheme, utilizing Homomorphic Encryption to enable healthcare practitioners to privately search and access patients' documents. The practicality and benefits of S3PHER are further validated through end-to-end deployment and use-case analyses, with tests on real datasets demonstrating promising execution times.

\end{abstract}
\section{Introduction}

Technology has experienced a tremendous evolution in the last few decades, as evidenced by the quality and complexity of consumer-available technological devices such as computers and smartphones. This evolution resulted not only in the widespread use of these devices in people's daily lives but also in an increase in service digitization across a wide range of industries.

Healthcare is one of the sectors that benefits most from these developments. The digitization and sharing of health data have well-known advantages on several levels. For example, personalized medical care and informed clinical decision-making are facilitated by the fact that healthcare professionals can access patients' electronic medical records ubiquitously ~\cite{kruse_use_2018}. 
On another level, the sharing of large amounts of health data can be used by researchers to train highly accurate Machine Learning predictive models. 
However, there are quite a few risks resulting from data sharing using electronic means. Some of the most private and sensitive information about a person can be found in healthcare data. This information includes personal identifiers like name, age, and address, as well as medical history, diagnosis, and treatment details. Since this data contains extremely private information, if compromised, it could have serious consequences such as financial fraud, identity theft, and even threats to one's safety~\cite{PAUL2023571}. For healthcare providers, data breaches can result in significant financial losses, and may negatively affect the patient trust, which can decrease the effectiveness of healthcare systems and have negative consequences for public health~\cite{Pool2024}.


To address these concerns, several organizations worldwide have developed legislative frameworks aimed at safeguarding patient information while fostering innovation and efficiency in healthcare delivery. Health Insurance Portability and Accountability Act\footnote{https://www.cdc.gov/phlp/publications/topic/hipaa.html} (HIPAA), valid in the United States of America, and the General Data Protection Regulation\footnote{https://data.europa.eu/eli/reg/2016/679/oj} (GDPR), valid in the European Union, are the two most known legislations that aim to tackle the issue of data protection.
In the healthcare sector, this requires organizations to obtain patient consent before processing health-related data and to report data breaches within a tight timeframe. However, several challenges still persist~\cite{RGPDdrawbacks}. For instance, usually, patients knowledge on who accesses their data and when the data was accessed is very limited, which is a severe issue.


%
With the introduction of cloud-based storage and data sharing platforms, among other things, the accessibility to health data has improved, as well as patient-centric data control and systems interoperability. This has led healthcare providers to rely heavily on Electronic Health Records (EHRs), telemedicine, and other digital tools. On the other hand, the risk of data breaches and unauthorized access to sensitive information also increases. Therefore, methods for protecting privacy and confidentiality must be continuously improved to ensure that sensitive data is never disclosed without prior authorization. New technological solutions must fulfil four key requirements, namely {Data Privacy and Confidentiality}, {Data Security}, {Data Integrity} and {Consent Management}, and failure to do so can have severe consequences~\cite{8888271}.
Simultaneously, maintaining the efficiency and practicality of these solutions is crucial~\cite{inproceedings11}.

 In this work, we focus on the problem of sharing health data between patients and healthcare practitioners, such as doctors, nurses, and technicians.  
 We want to ensure that patients' data is both private and accessible to practitioners from different organizations, when previously authorized by the patients. 
Several advancements and techniques have been developed such that these, and other security requirements, are ensured in the healthcare sector to preserve data privacy. They are classified, by some authors, as cryptographic and non-cryptographic~\cite{8726303}.
Most non-cryptographic approaches rely on access control policies such as Attribute-based Access Control (ABAC), and Identity Based Access Control (IBAC). These access control policies are essential as they define the limits of privileges and thus help ensure the privacy of information. However, by themselves, they do not offer complete security, since data is automatically exposed if these access control policies are maliciously overridden.
To tackle this problem, these control policies are commonly combined with encryption techniques~\cite{article111},  namely Symmetric Key Encryption (SKE) and Public Key Encryption (PKE), to ensure data privacy and confidentiality.
More suitable alternatives for modern applications include Searchable Encryption~(SE), which allows searches to be performed privately over an encrypted database, Proxy Re-Encryption~(PRE), which offers a secure method to delegate decryption rights between different users using different keys, and Homomorphic Encryption~(HE), due to its unique property of allowing computations to be performed over encrypted data.
Well known examples of combinations of cryptographic approaches with access control policies are Attribute Based Encryption~(ABE)~\cite{7118738} and Identity Based Encryption~(IBE)~\cite{IBEcrypt}.


\textbf{S3PHER} In this work, we present S3PHER, a new system for the sharing of health data between patients and healthcare practitioners. This system is designed for a context where the patient is provided with full control over who can access what data, and when. It adheres to data privacy regulations such as GPDR by respecting patient consent. In S3PHER, the patient is asked to consent whenever a practitioner requests access to a healthcare document. Only after providing this consent, the document is available for being accessed by that practitioner. 
Additionally, patients encrypt their healthcare documents before uploading them to a cloud storage service, referred to as a proxy, which stores them in encrypted form. It acts as an intermediary, requesting patient approval before granting access to healthcare practitioners. Patients retain the right to accept or reject access requests. The proxy itself cannot decrypt the data, however, it can re-encrypt the documents for healthcare practitioners who have valid consent, enabling them to decrypt it.  The system achieves end-to-end data privacy by employing a non-interactive PRE scheme~\cite{Umbral} that allows for the creation of a re-encryption key without the need for the healthcare practitioner's private key.

The ability to search patients' health data by keywords securely and privately is also provided by S3PHER. This functionality is enabled through a new SE scheme that uses HE techniques. The search process is done considering an index that is built by the patient before encrypting the health document. This approach allows healthcare practitioners to perform searches on encrypted indexes without exposing the actual content of the patient's data or the keywords they are searching for. It also ensures that neither the practitioners nor the cloud storage system ever sees the actual data or the keywords themselves. This separation is critical in maintaining the integrity of end-to-end privacy. As a result, searches can be conducted efficiently and securely, ensuring that sensitive health information remains confidential and accessible only to those with valid consent.




\noindent The main contributions of our work are the following:
%
\begin{enumerate}
\item A novel data-sharing system, S3PHER, designed for patients to control what healthcare data is accessed by whom, and when. This system ensures end-to-end privacy through encryption. It also allows the data users to search for documents privately.

\item 

A performance analysis of a Rust implementation of S3PHER that assesses the impact of the size of the healthcare documents on the execution time.

\item An an end-to-end evaluation by identifying different use cases and assessing the ability of S3PHER to fulfil their requirements. An estimation of the duration of each process in those use cases is also provided.
\end{enumerate}


\section{Background}
This section briefly introduces the three main concepts used throughout this work, namely Homomorphic Encryption, Proxy Re-Encryption and Searchable Encryption. 

\subsection{Homomorphic Encryption}

HE schemes are mainly categorised according to the number of operations allowed on the encrypted text. If the scheme allows for computations using only one operation an unlimited amount of times, it is a Partially Homomorphic Encryption (PHE) scheme.
Fully Homomorphic Encryption (FHE) schemes allow for more than one operation to be performed an unlimited amount of times. This type of scheme represents the pinnacle of HE schemes in terms of functionality. However its drawbacks are also much more severe than other less functional schemes, which causes FHE to have a much higher computational cost, and therefore worst performance than PHE. One of the most representative schemes, in the current landscape of FHE is TFHE~\cite{chillotti_tfhe_2020}, which 
is mainly built to be a fast FHE scheme, making it a very promising choice for several applications that could benefit from the functionalities of FHE but cannot afford the computational toll. Moreover, TFHE’s ability to encrypt bits individually also increases its flexibility in real use-cases.

Various libraries are now capable of providing implementations of the most used HE schemes, such as  HElib\footnote{https://github.com/homenc/HElib}, from IBM, which is the oldest HE library, and Microsoft SEAL\footnote{https://github.com/microsoft/SEAL}, introduced by Microsoft at the end of 2018 and that supports the same schemes as HElib. Other libraries are fully dedicated to specific encryption schemes, such as 
TFHE (Fully Homomorphic Encryption over Torus)\footnote{https://tfhe.github.io/tfhe/} for its respective scheme. In this work, we use the TFHE-rs\footnote{https://github.com/zama-ai/tfhe-rs} library, which was proposed by Zama\footnote{https://www.zama.ai/} as a further optimisation of the TFHE scheme.

\subsection{Proxy Re-Encryption}

PRE schemes usually involve three main entities, namely {Delegator}, {Delegatee} and {Proxy Server} (PS). To allow secure delegation of decryption rights between the delegator and delegatee, the proxy server uses a previously generated re-encryption key to transform the ciphertexts encrypted under the Delegator’s public key into a ciphertext that can be decrypted by Delegatees using their own private key. Khan~\cite{KF16} proposes a classification of PRE schemes, considering properties such as directionality, single/multi-use, interactivity, transitivity, verifiability and others. Although this primitive is generally applied in secure sharing of sensitive data or files, the choice of properties depends on the specific use case. Nuñez et al.~\cite{NUNEZ2017193} analyzed and compared several PRE schemes according to their properties, making it a reference work to consult when deciding which PRE scheme is more appropriate for a specific application.

In healthcare applications, some key properties have been identified as crucial~\cite{rodrigues.10.1007/978-3-031-54204-6_11}. These include non-interactivity, which states that the generation of the re-encryption key does not require the secret key of delegatees, and unidirectionality, meaning that re-encryption is only possible from delegator to delegatee and not vice versa. Slightly different from these two properties, but still essential, is the verifiability property, which allows the verification of re-encryption operations.

\subsection{Searchable Encryption}

Similar to PRE, SE is also a system that involves different entities and can be suited to different use cases. It enables searching through encrypted data without having to decrypt it at any intermediary stage. A standard SE has three main entities, namely {Data Owner} (DO), responsible for encrypting the sensitive data and outsourcing it to the cloud server, {Data User} (DU), responsible for querying DO's encrypted data, and {Cloud Server} (CS), which is responsible for storing the encrypted data and proving the SE service, i.e., searching the data and keeping search data structures updated.

The most interesting use cases for SE involve the use of a public CS, i.e., when sensitive data are outsourced to a CS. These cases require special care, since sensitive information will be available in the public cloud. SE schemes can be categorized as symmetric or asymmetric, similar to traditional encryption schemes, and can use one or more cryptographic techniques. They are mainly categorized based on how they perform the search protocol, i.e., via auxiliary index or via sequential scan of the whole document, how many users are involved, which functionalities are available when searching, and other miscellaneous properties~\cite{math11132948}. 
The search functionalities of a scheme can range from the single/multi keyword search duality tfo other functionalities such as fuzzy keyword search, ranked search, phrase search, and several others~\cite{article112}. The choice of properties is based on the specific use case. However, in the context of healthcare, a few properties can be highlighted due to their importance, such as index-based search, which uses an index to perform searches instead of searching directly on the sensitive data, and multi-keyword search, to allows search queries containing multiple keywords~\cite{article111}. Another characteristic extremely important for SE solutions in healthcare is to support dynamic databases, i.e., allow for insertion and deletion of files~\cite{article113}.



\section{S3PHER}
In this section we describe S3PHER, our system that allows secure and private sharing of healthcare data between  patients and their healthcare practitioners while enabling patient-centric data and consent management. 



%

There are three main entities in the proposed system, namely Data Owner (DO), {Data User (DU)} and {Proxy Server (PS)}. The DO represents the patient, responsible for encrypting their medical resources as well as generating and encrypting the metadata required to allow for encrypted search queries. These queries are generated and sent by DUs, represented by healthcare practitioners. The PS is a semi-trusted third-party hosted server responsible for managing almost every component needed in the system, such as public keys, documents, etc.


We assume that the hospital, or any other healthcare infrastructure, is responsible for managing the system, i.e., assuring that the system works as expected, both on the patient and the practitioner side. 

\subsection{Concept Idea} \label{concept}

\begin{figure}
    \centering    \includegraphics[width=0.6\linewidth]{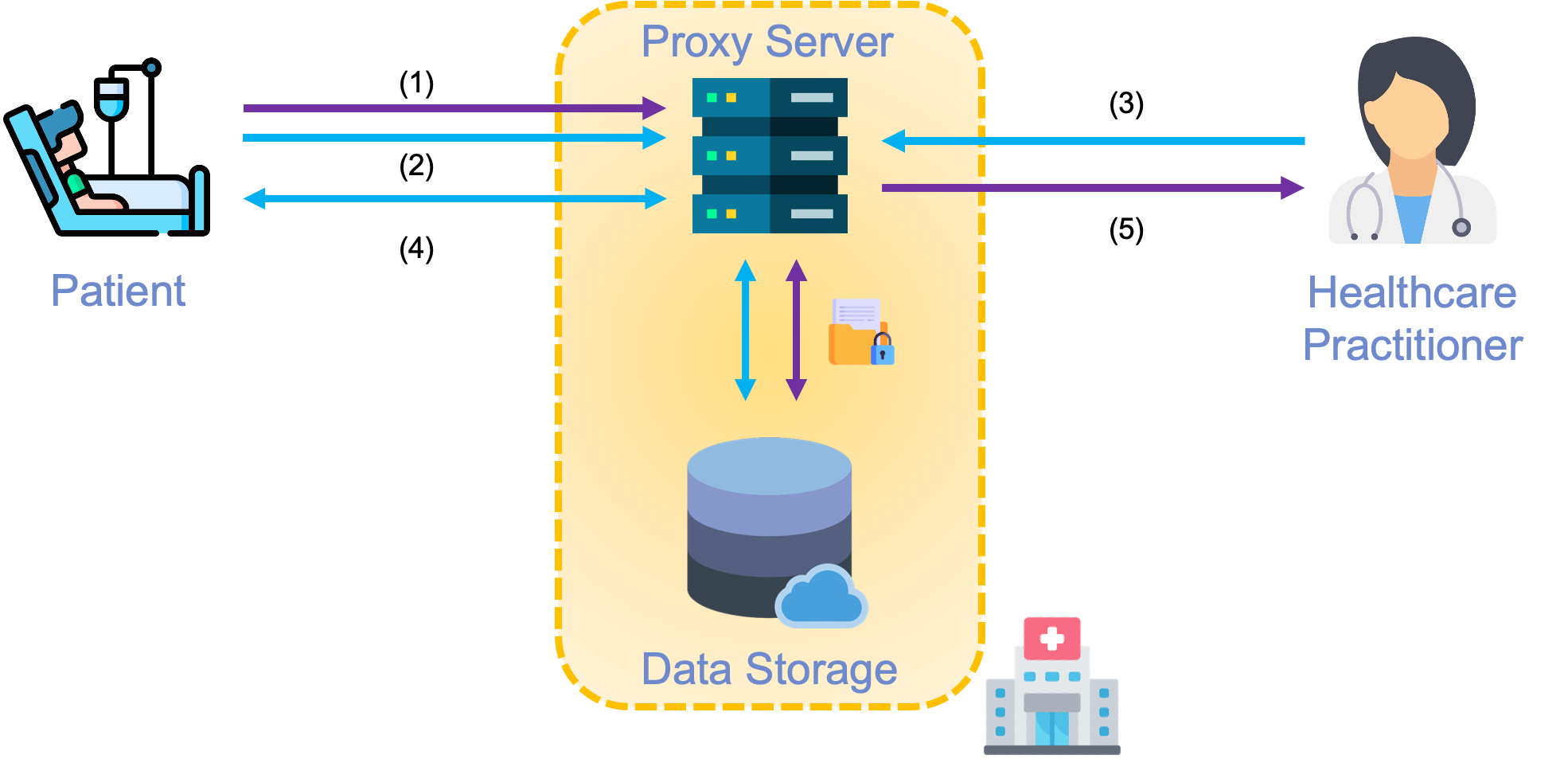}
    \caption{S3PHER System Design (Blue arrows represent steps related to the \textit{Searchable Encryption Module}; Purple arrows represent steps related to the \textit{Data Sharing Module}).}
    \label{diagramenti}
\end{figure}

S3PHER is composed of a \textit{Data Sharing Module} and a \textit{Searchable Encryption Module}, where the former addresses the processes directly related to the sharing of healthcare data, and the latter the related to the search process. Fig.~\ref{diagramenti} provides a high-level view of S3PHER system design.

Our \textit{Data sharing Module} follows an approach similar to the one proposed by Rodrigues et al.~\cite{rodrigues.10.1007/978-3-031-54204-6_11}, with the differences being highlighted later in this section.
We have used 
Umbral's PRE scheme to ensure the security of sensitive medical data. Umbral is a threshold PRE scheme proposed by Nuñez~\cite{nunez2017proxy} in 2017, which is unidirectional, non-interactive and verifiable. All these properties are crucial in PRE applied to the sharing of healthcare data.
The basic idea behind this scheme, is to employ the re-encryption techniques over the key, instead of over the ciphertexts, as the majority of PRE scheme do. This allows Umbral's scheme to leverage SKE while also using encapsulation techniques to re-encrypt the symmetric key. In other words, SKE can be used to ensure the security of sensitive data, while encapsulation techniques are applied to protect the key of the SKE scheme, at every stage in the re-encryption process. Two key steps of this module are represented in Fig.~\ref{diagramenti}, namely (1) and (5), representing the initial outsourcing of the encrypted healthcare files, on the patient side, and its final decryption, by the practitioner.

The \textit{Searchable Encryption Module} uses HE, more specifically a FHE scheme, namely TFHE~\cite{chillotti_tfhe_2020}. The reason for choosing HE relies on its unique ability to allow for computations over encrypted data while maintaining security and confidentiality at all times. Moreover, TFHE was a natural choice since it is, currently, one of the most promising FHE schemes~\cite{pedroAI10.1007/978-3-031-38318-2_36}, while also having several implementations publicly available. 
The idea for this module is to build an index, beforehand, to identify which subset of keywords properly describes each healthcare file. More specifically, we consider a matrix-like structure in which rows represent keywords, columns represent files and a given entry is either $1$ or $0$, depending on whether the respective healthcare document (column) contains the respective keyword (row) or not. Additionally, a column with each keyword encrypted is added as the left-most column in the matrix index. 

After building the index, a patient encrypts every entry using TFHE's public key and sends the encrypted result to the PS for storage, represented in Fig.~\ref{diagramenti} by (2). When the DU wishes to query the patient's resources (3), he/she chooses a keyword and encrypts it using the patient's TFHE public key.
Then, sends the result to the PS, which performs computations using the user's encrypted query and the patient's encrypted index (under the same key). The first computation performed involves the encrypted query and the left-most column containing the encrypted keywords, which results on an encrypted binary tuple with non-null entries if and only if a match was found. This tuple allows to generate a new version of the index matrix with every row zeroed except the row correspondent to the non-null entry of the tuple, which is the matched keyword. Finally, the PS applies an OR operation across every column of this new matrix, originating the result vector, which is an encrypted binary tuple with $1$s in the entries correspondent to the files described by the queried keywords.
Note that the PS will not be aware if a match was found or not, or where it was found, at any stage of the computation process. The encrypted result must be sent to the patient, who uses his/her TFHE secret key to decrypt it and check which resources have been requested. Finally, the patient either authorizes or declines access to the requested resources (4), and informs the PS of his/her decision, concluding the Search protocol. 
Note that when the Search protocol is concluded, the Data sharing protocol is meant to be initiated. In comparison with Rodrigues et al. proposal~\cite{rodrigues.10.1007/978-3-031-54204-6_11}, where the PS would receive a request from the DU containing the resources IDs wished to be retrieved, in our proposal, we avoid giving such information to DUs. This is achieved by the DO informing the PS which resources, from those requested via keyword search, are authorized for retrieval.

\subsection{Design Goals}


 The main focus in the design of S3PHER was to avoid as many computations as possible on the patient and practitioners side. To this end, the PS is encharged with every task that does not require the intervention of any of the other entities, specially storage and computation tasks. This allows any resource server(s) in the system to be exclusively dedicated to help clients in tasks that cannot be delegated to the proxy without compromising the client's privacy. This is not the case in Rodrigues et al. proposal, since the majority of tasks require some interaction between entities or/and components.
 
The running time performance of S3PHER was also a major concern. To address this, the implementation of S3PHER was carried out on the Rust execution environment, which provides a fast implementation of TFHE. Moreover, an appropriate Umbral-PRE Rust implementation was used as well. 




\textbf{Threat Model} \label{threatm}
Since our system grants the patient total control over their sensitive data, we design a threat model that considers all threats from either a semi-honest PS or a malicious user (healthcare practitioner).%
The system must ensure that no sensitive information is leaked to any involved entity that does not have permission to access it.

We consider a \textbf{Semi-honest PS}, that although it does not intend to maliciously interfere in the protocol, is trying to learn as much as possible from all available data. More specifically, a semi-honest proxy is trying to learn information about a patient's encrypted medical files as well as the respective encrypted metadata. Moreover, it may also attempt to extract information from keyword queries performed by the user. 
Even if some information is leaked to the PS, it should not be able to access it or, if the information is in the plaintext format, it should not reveal any relevant information about sensitive data.


A \textbf{Malicious Practitioner} might attempt to impersonate another practitioner and query sensitive data from patients. They can also conduct unjustified queries on their patients to access more sensitive information, which could then be used for personal gain.
These adversaries should not be capable of accessing a patient's sensitive data in either scenario.

Not all threats to S3PHER originate from the entities involved in the system. We can also have a \textbf{Malicious External Actor} that may attempt to disrupt the process during communication.
The system must ensure that even if this adversary can extract any data between communication rounds, it will not be able to access any sensitive data since it does not have valid consent.


The security analysis will be performed per adversary, since we assume that they do not collude.
A system that proves to be secure against these adversaries ensures the confidentiality and privacy of patients' sensitive data.

\subsection{Security Analysis} \label{secana}

In this subsection, we discuss the security of our system, according to the threat model defined in Section~\ref{threatm}. Three main adversaries were identified in the threat model, namely, Semi-honest proxy, Malicious practitioner, and Malicious External Actor. To show that our scheme fulfils the security requirements, adapted to the proposed use case, we discuss, in detail, the risks that each adversary brings. In this way, we can show that sensitive data is never at risk of being exposed to unauthorized parties.

\textbf{Semi-honest PS}
It is not a malicious entity, therefore it does not present the same threat as a malicious healthcare practitioner. However, we wish to prevent it from learning anything that it is not meant to learn.

The PS has access, in plaintext format, to the number of files a patient owns, the status of sharing requests, identifier of the DU requesting a file and identifier of the owner of the file, to correctly perform its tasks.
The remaining data, involved in the protocol of S3PHER, is either encrypted under TFHE, or under Umbral's PRE. If encrypted under TFHE, then the holder of the private key (patient) is the only one able to decrypt such ciphertexts. However, the patient never has access to ciphertexts encrypted by practitioners. On the other hand, if encrypted under Umbral’s PRE, either by the patient or practitioners, these ciphertexts can only be decrypted by the holder of the secret key.

\textbf{Malicious Practitioner}
This entity represents the biggest threat to our system, since it might be hard to distinguish a malicious healthcare practitioner's query from any unusual queries performed by honest practitioners. This adversary can either be a user impersonating a patient's healthcare practitioner, or it can be an actual practitioner who developed malicious intentions. Regardless of the scenario, this adversary only has access to the patient's TFHE public key, which prevents him from decrypting any data whatsoever, unless the DO provides such consent. This, however, does not have to be the case due to the revoke functionality, since in the event of a suspicious query by any user, the patient can opt to instantly decline that request.

\textbf{Malicious External Actor}
This threat was introduced to highlight our use of cryptographic primitives in each step of the process. This assures that if any data is leaked at any moment of the communication process, an attacker would not be unable to decrypt or take any conclusions from such data since the same is end-to-end encrypted.

\section{Performance Evaluation}

In this section, we describe the tests performed to evaluate S3PHER and present the results obtained. The system's performance was evaluated regarding its running time. Both the Data sharing and Searchable Encryption modules of S3PHER are assessed.


The tests were performed using a device with the following specifications: CPU -- Ryzen 7 5700X (8 cores, 16 threads); RAM -- 32GB DDR4 2133MHz; OS-- Windows 11 23H2.
For each combination of parameters, the tests were performed 15 times, and we present the results in a table format that showcases the average across these 15 runs with the corresponding standard deviation. These tables can be found in Appendix~\ref{appendix}.

\subsection{Data Sharing Module}

The objective of this experiment was to assess how the running time of the different processes of the Data Sharing module varies according to different sizes of healthcare documents. For this purpose, randomly generated files of sizes 1MiB, 10MiB, 100MiB, 1GiB and 10GiB were used to assess different stages of the data sharing protocol implementation.
The results obtained are presented in Table~\ref{PREtime}, on Appendix~\ref{appendix}.

Observing the results, we concluded that none of the parameters related to Umbral's PRE scheme is significantly impacted by the size of the document. On the other hand, both encryption and decryption are affected by the document size. However, the growth rate for the running time is smaller when compared to the growth in the size of the documents. Moreover, although encryption appears to be more time efficient than decryption when looking at the light-sized documents, it can be seen that in larger files, the scenario inverts, and the growth rate of encryption appears to be much greater than the one seen in decryption. 
Another note on the encryption results is its high standard value, which means that this stage suffers a big variation. 
Finally, the global analysis consists of a timer set at the beginning of the process until the process is concluded. Note that this global analysis gives an upper bound on the actual time needed to perform this set of tasks.


   
   











						

\subsection{Searchable Encryption Module}
Since the search functionality works almost independently of the documents, excluding the initial keyword extraction, there is no interest in testing this functionality over documents of different sizes. Due to how the search is implemented, i.e., across all rows and columns of the File keyword matrix, we can estimate the complexity of the components of this module to be similar to the dimension of the matrix. Therefore, the tests have, as parameters, the number of keywords and the number of files, which are represented by the rows and columns, respectively.
For this test the parameter "number of keywords" is chosen from the set \{1,4,8,16,32,63\} and the parameter "number of files" from \{1,8,32,64,128\}.

This test was meant to analyse the running time of the two main components of the search module, namely ``Matrix generation'' and ``Search''. Table~\ref{matrixnsearchtime}, on Appendix~\ref{appendix}, shows the results obtained.

As expected, the running time tends to increase as the matrix's dimension increases, however, we can observe that the number of keywords has much more impact on the running time than the number of files.
Nonetheless, even for the highest dimensional case, the matrix generation is still very efficient.

When analysing the search running time results, the first thing that stands out is the difference compared to the matrix generation results. This is, however, not surprising, since the search process involves much more computations and interactions between entities.
On the other hand, similarly to the previous table, the search running time also increases as the dimension increases. Moreover, we can also observe that the number of keywords has much more impact than the number of files, similar to the matrix generation. This impact is, however, more noticeable in the search protocol, mainly in the comparison process needed to find keyword matches.



    











\section{Healthcare System Simulation} \label{usecasesap}
In this section we describe, explicitly, a few healthcare use-cases in which our proposed system could be advantageous. Moreover, we performed simulations on each of these use-cases, in order to evaluate our system's end-to-end performance.

\subsection{Simulation Set-up}
Firstly, we define the healthcare environment considered for the simulation. Up until this point, we considered only one patient, however, this in not the case in a real-world scenario. Nonetheless, the key protocols remain unchanged in a multi-patient scenario, since we expect healthcare practitioners to perform keyword searches over a specific patient's medical data at a time. Moreover, this choice of patient consists in a plaintext search over all patient's IDs, which is not expected to have a costly impact on the S3PHER's performance.

In our simulation, we considered a healthcare system that is composed by 975 patients, and a total of 13372 files, randomly distributed across all patients, such that every patient had at least 1 file. This resulted in an average 13,71 files per patient.
To construct this system, we used three different datasets, namely Kvasir~\cite{10.1145/3193289}, CovidX~\cite{Gunraj2022} and Facial Emotion Recognition(FER 2013)~\cite{challenges-in-representation-learning-facial-expression-recognition-challenge}. Kvasir is a collection of annotated medical images from the inside of the Gastrointestinal tract. CovidX is a collection of over 13000 X-Rays, with annotations related to the Covid-19 diagnosis. FER is a dataset that contains images of faces displaying 7 different emotions. Although the latter dataset it is not directly related with healthcare, it was also chosen to add more realism to the simulation, as it is possible that in a real-world scenario, where it may occur the sharing of non-medical files, e.g. profile pictures.
To guarantee a map between keywords and files, an initial pre-processing on the datasets had to be done. The use of Kvasir and FER was straightforward, while CovidX required a pre-selection of used features (to be used as keywords). This resulted in a total of 32 considered keywords, 14 from CovidX, 8 from Kvasir and 10 from FER.


\subsection{Use-cases}
The main goal of S3PHER is to provide a secure and efficient method for healthcare practitioners to request and access a patient's medical data, as long as the patient grants consent. However, more specific use cases, simulating real life scenarios, can be derived from the stated goal, which is our intention in this section.
We consider three main use-cases, namely \textbf{Doctor's appointment}, \textbf{Medical Report} and \textbf{Doctor Change}.

\textbf{Doctor's Appointment}
This use-case represents a simple doctor's appointment, in which the doctor needs to consult some of the patient's medical files.
Suppose a patient, named Samantha, schedules a doctor's appointment, either online or in-person, and, in the middle of the appointment, the doctor needs to consult some of Samantha's X-ray exams. Using our system, the doctor logs in, and builds their query over Samantha's sensitive data, using the following keyword query: "Pneumonia OR COVID-19".
After the search has been processed, the doctor informs Samantha, who accesses our system on her smartphone, to review the doctor's request and accept it.
After the sharing is complete, the doctor can download and access the files on their computer and perform Samantha's diagnosis as accurately as possible.

In our simulation, "Samantha" was chosen at random, which resulted in user id "CP\_1365", which has 20 files.
Considering the obtained results, we estimate this use-case to take at most 3 minutes to be concluded. This time is estimated by assuming Samantha takes around 1 minute to access the system on their phone and the Doctor takes close to 1 minute to build their query. The remaining time is divided into the search protocol, accepting the search request and downloading the results. The search protocol took around 32 seconds to be concluded, while processing the accepted request took only 14 seconds and downloading plus decrypting the file name took a little over 5 seconds. Finally, the download of the result files depends on the size of the files, and since our databases are constituted by images, the download of the resultant files, which were 7, took 10 seconds. To summarise, Samantha and her doctor are involved in the protocol for the following time,
\begin{itemize}
    \item Samantha - 1:14 minutes (Accessing app + process accepted request)
    \item Doctor - 3 minutes (Build query + search process + wait for permission + downloads).
\end{itemize}

Note that the duration of making a search request is not significantly impacted by the patient's number of files. However, reviewing and accepting it, can only be done after the search has been concluded. The duration of this latter task is dependent on the number of files of the patient, with a number of files below 10 adding just milliseconds to the processing time, while a number of files around 60, can add up to a minute, considering a single-keyword query.


\textbf{Medical Report}
Suppose that after the Doctor's appointment depicted in the previous use-case, Samantha was told to do some extra exams so the doctor could better evaluate her well-being and proceed to write an appropriate report on her health condition. The doctor instructed Samantha, that when the results were available, in order to avoid scheduling an appointment just to show the results, she would perform a keyword query over Samantha's medical files, and all she has to do is review the request and accept it, on her own smartphone. This use-case aims to show that our system does not rely on physical interaction between entities.
Since this use-case does not require Samantha and her doctor to be in an interaction, its harder to estimate the time that each of the participants is going to take until accessing the system and seeing a pending request, in Samantha's case, or a pending download in the Doctor's case. Nevertheless, we summarise the participants' time cost as follows,

\begin{itemize}
    \item Samantha - 1:14 minutes + $\epsilon$
    \item Doctor - 3 minutes + $\epsilon$ + $\delta$,
\end{itemize}
where $\epsilon$ is the time that Samantha takes to log into the system and see a pending request, and $\delta$ is the time that Samantha's doctor takes to log into the app and see that her request has been accepted. As we can easily see, the difference in time cost between this use-case and the previous one differs only in the time it takes from both Samantha and her doctor to acknowledge each others actions. In the previous use-case we assume it's instant, which is not the case here.

\textbf{Changing Doctors}
This use-case intends to highlight the revoke functionality, which is a key component in our system, since it easily allows any doctor to request and access medical files from any patient registered into our system.
Suppose Samantha is no longer happy with her current doctor and she decides to change to another doctor within the same medical infrastructure. To do this, she can simply access our system and revoke access to all files shared with her current doctor, such that he has no longer access to them. 
We tested this use case by revoking the access to the 7 files returned in the previous use-case's queries. This took a total of 19 seconds to conclude.

\textbf{Malign Access Attempt}
To conclude this section, we present a misuse, which derives from the use-case \textit{Medical Exams}.
Suppose Samantha's doctor has malicious intentions and performs a query on Samantha's medical files in order to use them for personal gain. We simulated this misuse by performing a different query from the previous use-cases, specifically, "NOT happy", which is an odd query given Samantha's previous history with her doctor. The query takes around 16 seconds to be processed.
After $\epsilon$ time has passed, Samantha accesses our system on her smartphone and finds out that her doctor made a unexpected request. Suspicious of her doctor's request, Samantha declines it promptly, taking only 1.8 seconds to do so, and proceeds to takes immediate action by reporting the incident to the clinic's administration. Moreover, if she wishes, she can revoke her doctor's access to any medical files previously shared.
Note that, it can be unclear if a request is suspicious or not. However, the patient has the option to either decline or accept, allowing for extra time to make a more informed decision.



\section{Related Works}


In this section, we discuss how S3PHER fits into the literature, highlighting the main advantages that our system has when compared to existing solutions.

\subsection{Proxy Re-Encryption and Search}
In 2015, Liang et al.~\cite{7118738} proposed a searchable Attribute-Based PRE (ABPRE), supporting keyword searches, which can be applied to various use-cases, including in the healthcare sector. Their proposal uses ABPRE to allow secure data sharing and uses Attribute-Based Keyword Search (ABKS) to allow for keyword searches in the system. Combining these techniques, the system can perform search queries through any owner's ciphertexts, after performing the proper re-encryption. Although their proposal gives the patient some granular access control, it does not give the patients total control over their data, since any user with the correct set of attributes will be able to search and access sensitive data.


In 2021, Niu et al.~\cite{WOS:000608044300042} proposed a Eletronic Medical Record (EMR) data-sharing scheme supporting multi-keyword search. Their proposal is a simplification of Chen et al.~\cite{10.1007/978-3-319-11698-3_7} work applied to two healthcare use-cases. In their system, when the DU performs an encrypted search, an authorization request is sent to the DO which has to provide consent for the  PRE scheme to be used to share the EMR securely, and generate the query token, which is then submitted for computations.


An year later, in 2022, Niu et al.~\cite{WOS:000829745700005} proposed a similar system. However, relying on two servers, namely Proxy server and Medical server, where the former is responsible for storing the search index, and the latter is responsible for storing the encrypted EHR. Although this proposal achieves fine grained access control similarly to our scheme, it does not provide a revoke functionality, which prevents DOs from managing any previously given consent.

Contrary to these works, in our proposal, we chose HE as the building block of our SE scheme, due to its unique ability to allow computations to be performed over encrypted data, which allows our system to have more functionalities while ensuring end-to-end data privacy and confidentiality. Moreover, our system comprises an authorisation protocol that gives the patients a sovereign control over what data can be accessed, by whom, and when. This allows a better fine-grained access control than solutions relying on ABKS.


\subsection{Homomorphic encryption and Search}

In 2020, Guo et al.~\cite{GUO2020102021} proposed Privacy-preserving Image Search (PPIS), which consists in a Convolutional Neural Network framework that allows the classification and searching of large-scale encrypted medical images. 
The encryption of the medical images was performed using the Paillier cryptosystem~\cite{paillier_public-key_1999} to allow the operation addition on encrypted data. However, this proposal lacks any sort of access control.

In 2021, Li et al.~\cite{LI2020101701} proposed a self-serviced medical diagnosis scheme based on secure multi-party computation. Their scheme allows a patient to query a hospital's disease database with his/her encrypted medical data, and receive the suited treatment for any diseases that were found in the search phase. HE is used to compute similarity scores, outputting the disease that the respective patient is most likely to have. This proposal uses an access control strategy that relies on signature keys and timestamps, such that every time patients want to consult their data, a signature key must be provided, and the timestamp must be valid. On the other hand, their proposal does not allow patients to revoke access from users who had been previously authorised, which might prevent patients from having total control over their sensitive data.

Also in 2021, Tahir et al.~\cite{9541339} proposed a novel architecture aimed at efficiently and effectively trace COVID-19 contacts. Their proposal uses blockchain network for storage, and relies on HE to encrypt the data meant to be searched for at a later stage. The search is performed to find all contacts of a patient that is suspected of being positive for COVID-19. These data are decrypted by healthcare practitioners or governmental entities, which must have been given consent by the users, however, once again, this proposal lacks a revokability functionality, in order to ensure total control to the patient.


In 2022, Ali et al.~\cite{WOS:000753704500001} proposed a SE scheme leveraging blockchain and HE. The latter is used as the foundation for the search process while Blockchain is responsible for ensuring security and efficiency. Their proposal allows a DU to perform multi-keyword queries over a patient's data, as long as the DU provides access control requirements. Their access control strategy combines ABE, anonymity, and trust value, where the latter allows DOs to set trust values for DUs.
In the same year, Iqbal et al.~\cite{s22124432} proposed a system that allows for secure searches over encrypted audio data that has been outsourced to the CS in a medical context. Their approach involves using the BGV FHE scheme to encrypt the data files, which are then sent to the CS for storage. To perform a search, users must have a valid consent and the authorisation keys granted by the DO.

Contrary to most proposals combining PRE and SE, these systems appear to be less concerned with access control policies, which is a property inherent in to the great majority of PRE schemes. Moreover, to the extent of our knowledge, there are only three proposed works that explicitly combine PRE, SE and HE~\cite{XIE202212,10235914,10149415}. However, neither of these approaches was designed to fulfil the security requirements of data sharing in the healthcare sector.
Differently from the works above, S3PHER combines PRE, SE and HE to tackle the main issues of the problem of data sharing in healthcare. Our system gives patients total control over their healthcare data, while also allowing key functionalities such as search over encrypted data, using SE and HE, and a secure and efficient method to share encrypted healthcare data using PRE.

\section{Conclusion}
In this paper, we proposed a system that combines PRE, SE and HE, to allow efficient search queries over encrypted data as well as secure sharing of healthcare data. TFHE is used as the foundation of our SE scheme, since every index constructed, as well as queries, are encrypted with this cryptographic primitive. Umbral's PRE scheme is used to ensure efficient and secure sharing of medical files between a patient and any of their health practitioners. Contrary to other state-of-the-art proposals, we do not implement a multi-key feature, and the reason is that the problem we are trying to solve does not require a practitioner to operate between ciphertexts of different patients. Nonetheless, such a feature could be easily achieved using our PS.
The proposed system was implemented using the Rust variants of Umbral and TFHE, and several tests were performed to analyse both running time and the size of relevant elements. The results show that besides the parameter number of keywords, all other parameters do not greatly impact either running time.
Lastly, we discussed and analysed specific healthcare use-cases in which our system could be a great advantage, both security, and performance-wise. The set-up for this simulation consisted of several patients, each with an arbitrary number of files, obtained from real-world databases containing the most common file formats.

Given the current status of S3PHER, and its main purpose, we aim to further extend S3PHER's privacy and confidentiality assurances. More specifically, we highlight two main problems that we seek to solve shortly, namely the compatibility of S3PHER with Fast Healthcare Interoperability Resources, which is a standard format for data sharing in the healthcare sector, and the ability to perform audit trails, which may be provided using Blockchain technology.

\begin{credits}
\subsubsection{\ackname} 
This work has been supported by the UIDB/00760/2020 and UIDP/00760/2020 projects.
\subsubsection{\discintname}
The authors have no competing interests to declare that are
relevant to the content of this article.
\end{credits}
%
%
%
\newpage
\bibliographystyle{splncs04}
\bibliography{mybibliography}
\appendix 

\newpage

\section{Experimental Results} 
\label{appendix}

To analyze the performance of Umbral’s PRE scheme within S3PHER, tests were conducted over its main stages, namely Encryption and Decryption. These tests analyze the impact of file size in these phases. Also a Global perspective that represents the whole process, i.e., from the beginning of Umbral’s PRE protocol until it is concluded, is considered. The results are presented in Table~\ref{PREtime}.

\vspace{-0.5cm}

\begin{table}
    \centering
    \resizebox{8cm}{!}{
\begin{tabular}{ccccccc}
    \toprule
     & \multicolumn{2}{c}{Encryption} & \multicolumn{2}{c}{Decryption} & \multicolumn{2}{c}{Global} \\
    
    \midrule
   File Sizes  & Average & StdDev  & Average & StdDev  & Average & StdDev \\
   
   \midrule
   
   1MiB     &   0.0017738	& 0.0002988 &   0.0168947 &	0.0001533  &   0.0386500	& 0.0006682    \\

    10MiB       &   0.0131566	& 0.0027700  &   0.0258313 &	0.0020094  &   0.0706933	& 0.0054992    \\

     100MiB   &   0.1622563	& 0.0422988  &   0.1106990 &	0.0149935    &   0.4309532	& 0.0687121    \\

     1GiB      &   1.1889699	& 0.0232585  &   0.8690198 &	0.0167867  &   3.3552095	& 0.0327308    \\

      10GiB    &   83.4834017	& 10.2941911 &   11.2849186 &	0.1886814  &   111.3354066	& 10.3355311   \\  

    \bottomrule

\end{tabular}
}
    \caption{PRE running time per stage.}
    \label{PREtime}

\end{table}

   
   











									




















\vspace{-0.5cm}

Table~\ref{matrixnsearchtime} showcases the running times obtained in the test performed to analyze the time cost of generating the index matrix (Matrix Gen), as well as performing the whole search process (Search).
These tests were performed varying the values of the considered parameters, namely Nº Keywords and Nº Files. These parameters dictate the dimension of the index matrix, which is a central component in both protocols.

\vspace{-0.6cm}

\begin{table}
    \centering
    \resizebox{7cm}{!}{
    \begin{tabular}{ccccccc}
    \toprule
    & &  \multicolumn{2}{c}{Matrix Gen} & & \multicolumn{2}{c}{Search} \\
    \midrule
   Nº Keywords & Nº Files & Average & StdDev & \hspace{0.5cm} & Average & StdDev \\
   
   \midrule
   
   \multirow{5}{*}{1} & 1 & 0.0000085 & 0.0000071 & \hspace{0.5cm} & 0.1135911 & 0.0425694 \\

   & 8 & 0.0000107 &	0.0000033 & \hspace{0.5cm} & 0.1630754	& 0.0087801 \\

    & 32 & 0.0000400 & 0.0000267 & \hspace{0.5cm} & 0.3703381	& 0.0155974 \\

     & 64 & 0.0000263 &	0.0000125 & \hspace{0.5cm} & 0.6590039	& 0.0446913 \\

     & 128 & 0.0000274	& 0.0000054 & \hspace{0.5cm} & 1.1945963	& 0.0691590 \\

     \midrule

     \multirow{5}{*}{4} & 1 & 0.0000286	& 0.0000138 & \hspace{0.5cm} & 0.4692102 &	0.0395151 \\

     & 8 & 0.0000392	& 0.0000141 & \hspace{0.5cm} &  1.0757792	& 0.0815401 \\

    & 32 & 0.0000410	& 0.0000089 & \hspace{0.5cm} & 3.1449836	& 0.1871842 \\

     & 64 & 0.0000503	& 0.0000068 & \hspace{0.5cm} &  5.8184655	& 0.2392645 \\

     & 128 & 0.0001035	& 0.0000334 & \hspace{0.5cm} & 11.1656595	& 0.2763816 \\

     \midrule

     \multirow{5}{*}{8} & 1 & 0.0000582	& 0.0000098 & \hspace{0.5cm} & 0.9210879	& 0.0124797  \\

     & 8 & 0.0000694	& 0.0000155 & \hspace{0.5cm} & 2.2364570	& 0.0949350 \\

     & 32 & 0.0000773	& 0.0000105 & \hspace{0.5cm} & 6.6863189	& 0.1598003 \\

     & 64 & 0.0000978	& 0.0000126 & \hspace{0.5cm} & 12.7394045	& 0.3698099 \\

     & 128 & 0.0001914	& 0.0000306 & \hspace{0.5cm} & 24.7864664	& 0.6486616 \\

     \midrule

     \multirow{5}{*}{16} & 1 & 0.0000926	& 0.0000102 & \hspace{0.5cm} & 1.8807355	& 0.0566454 \\

     & 8 & 0.0001302	& 0.0000169 & \hspace{0.5cm} & 4.6218205	& 0.1400157 \\

     & 32 & 0.0001915	& 0.0000267 & \hspace{0.5cm} & 14.0084332	& 0.3566607 \\

     & 64 & 0.0002383	& 0.0000159 & \hspace{0.5cm} & 26.9596536	& 1.2346072 \\

     & 128 & 0.0004121	& 0.0000942 & \hspace{0.5cm} & 51.7502386	& 1.4135845 \\

    \midrule

    \multirow{5}{*}{32} & 1 & 0.0001874	& 0.0000197 & \hspace{0.5cm} & 3.7984689	& 0.1062680 \\

     & 8 & 0.0002811	& 0.0000575 & \hspace{0.5cm} & 9.4172845	& 0.3779081 \\

     & 32 & 0.0003460	& 0.0000220 & \hspace{0.5cm} & 28.8329438	& 0.9610260 \\

     & 64 & 0.0004726	& 0.0000393 & \hspace{0.5cm} & 54.9305971	& 1.9372611 \\

     & 128 & 0.0008204	& 0.0003290 & \hspace{0.5cm} & 107.1033151	& 3.7332988 \\

     \midrule

     \multirow{5}{*}{63} & 1 & 0.0004724	& 0.0000513 & \hspace{0.5cm} & 7.5399973 &	0.2538821 \\

     & 8 & 0.0005677	& 0.0000407 & \hspace{0.5cm} & 18.7667887	& 0.7236851 \\

     & 32 & 0.0007543	& 0.0000734 & \hspace{0.5cm} & 57.2512293	& 1.7110481 \\

     & 64 & 0.0009233	& 0.0001178 & \hspace{0.5cm} & 107.5793417	& 2.1456864 \\

     & 128 & 0.0015773	& 0.0001740 & \hspace{0.5cm} & 210.3645677	& 4.5028557 \\

    \bottomrule
\end{tabular}    
}
    \caption{Matrix Generation and Search running times.}
    \label{matrixnsearchtime}

\end{table}





\end{document}